\documentstyle{mn}
\newread\epsffilein    
\newif\ifepsffileok    
\newif\ifepsfbbfound   
\newif\ifepsfverbose   
\newdimen\epsfxsize    
\newdimen\epsfysize    
\newdimen\epsftsize    
\newdimen\epsfrsize    
\newdimen\epsftmp      
\newdimen\pspoints     
\def\epsfbox#1{\global\def\epsfllx{72}\global\def\epsflly{72}%
   \global\def\epsfurx{540}\global\def\epsfury{720}%
   \def\lbracket{[}\def\testit{#1}\ifx\testit\lbracket
   \let\next=\epsfgetlitbb\else\let\next=\epsfnormal\fi\next{#1}}%
\def\epsfgetlitbb#1#2 #3 #4 #5]#6{\epsfgrab #2 #3 #4 #5 .\\%
   \epsfsetgraph{#6}}%
\def\epsfnormal#1{\epsfgetbb{#1}\epsfsetgraph{#1}}%
\def\epsfgetbb#1{%
\openin\epsffilein=#1
\ifeof\epsffilein\errmessage{I couldn't open #1, will ignore it}\else
   {\epsffileoktrue \chardef\other=12
    \def\do##1{\catcode`##1=\other}\dospecials \catcode`\ =10
    \loop
       \read\epsffilein to \epsffileline
       \ifeof\epsffilein\epsffileokfalse\else
          \expandafter\epsfaux\epsffileline:. \\%
       \fi
   \ifepsffileok\repeat
   \ifepsfbbfound\else
    \ifepsfverbose\message{No bounding box comment in #1; using defaults}\fi\fi
   }\closein\epsffilein\fi}%
\def\epsfsetgraph#1{%
   \epsfrsize=\epsfury\pspoints
   \advance\epsfrsize by-\epsflly\pspoints
   \epsftsize=\epsfurx\pspoints
   \advance\epsftsize by-\epsfllx\pspoints
   \epsfxsize\epsfsize\epsftsize\epsfrsize
   \ifnum\epsfxsize=0 \ifnum\epsfysize=0
      \epsfxsize=\epsftsize \epsfysize=\epsfrsize
     \else\epsftmp=\epsftsize \divide\epsftmp\epsfrsize
       \epsfxsize=\epsfysize \multiply\epsfxsize\epsftmp
       \multiply\epsftmp\epsfrsize \advance\epsftsize-\epsftmp
       \epsftmp=\epsfysize
       \loop \advance\epsftsize\epsftsize \divide\epsftmp 2
       \ifnum\epsftmp>0
          \ifnum\epsftsize<\epsfrsize\else
             \advance\epsftsize-\epsfrsize \advance\epsfxsize\epsftmp \fi
       \repeat
     \fi
   \else\epsftmp=\epsfrsize \divide\epsftmp\epsftsize
     \epsfysize=\epsfxsize \multiply\epsfysize\epsftmp   
     \multiply\epsftmp\epsftsize \advance\epsfrsize-\epsftmp
     \epsftmp=\epsfxsize
     \loop \advance\epsfrsize\epsfrsize \divide\epsftmp 2
     \ifnum\epsftmp>0
        \ifnum\epsfrsize<\epsftsize\else
           \advance\epsfrsize-\epsftsize \advance\epsfysize\epsftmp \fi
     \repeat     
   \fi
   \ifepsfverbose\message{#1: width=\the\epsfxsize, height=\the\epsfysize}\fi
   \epsftmp=10\epsfxsize \divide\epsftmp\pspoints
   \newcount\figskipcount
      \message{#1 \the\epsfysize  }
   \vbox to\epsfysize{\vfil\hbox to\epsfxsize{%
      \includegraphics{#1}%
      \hfil}}%
\epsfxsize=0pt\epsfysize=0pt}%

{\catcode`\%=12 \global\let\epsfpercent=
\long\def\epsfaux#1#2:#3\\{\ifx#1\epsfpercent
   \def\testit{#2}\ifx\testit\epsfbblit
      \epsfgrab #3 . . . \\%
      \epsffileokfalse
      \global\epsfbbfoundtrue
   \fi\else\ifx#1\par\else\epsffileokfalse\fi\fi}%
\def\epsfgrab #1 #2 #3 #4 #5\\{%
   \global\def\epsfllx{#1}\ifx\epsfllx\empty
      \epsfgrab #2 #3 #4 #5 .\\\else
   \global\def\epsflly{#2}%
   \global\def\epsfurx{#3}\global\def\epsfury{#4}\fi}%
\def\epsfsize#1#2{\epsfxsize}
\hyphenation{english}

\newif\ifAMStwofonts

\def\simgt{\hbox{\rlap{\raise 0.425ex\hbox{$>$}}\lower 0.65ex\hbox{$\sim$}}}
\def\simlt{\hbox{\rlap{\raise 0.425ex\hbox{$<$}}\lower 0.65ex\hbox{$\sim$}}}

\def\degree{^\circ}

\def\bj {b_{\rm J}}
\def\lb {L_{\rm B}}
\def\mb {M_{\rm B}}

\ifoldfss
  \ifCUPmtlplainloaded \else
    \NewTextAlphabet{textbfit} {cmbxti10} {}
    \NewTextAlphabet{textbfss} {cmssbx10} {}
    \NewMathAlphabet{mathbfit} {cmbxti10} {} 
    \NewMathAlphabet{mathbfss} {cmssbx10} {} 
  \fi
  \ifAMStwofonts
    \ifCUPmtlplainloaded \else
      \NewSymbolFont{upmath} {eurm10}
      \NewSymbolFont{AMSa} {msam10}
      \NewMathSymbol{\upi}     {0}{upmath}{19}
      \NewMathSymbol{\umu}     {0}{upmath}{16}
      \NewMathSymbol{\upartial}{0}{upmath}{40}
      \NewMathSymbol{\leqslant}{3}{AMSa}{36}
      \NewMathSymbol{\geqslant}{3}{AMSa}{3E}

      \let\leq=\leqslant 
       \let\ge=\geqslant
    \fi
  \fi
\fi 

\ifnfssone
  \newmathalphabet{\mathit}
  \addtoversion{normal}{\mathit}{cmr}{m}{it}
  \addtoversion{bold}{\mathit}{cmr}{bx}{it}
  \newmathalphabet{\mathbfit} 
  \addtoversion{normal}{\mathbfit}{cmr}{bx}{it}
  \addtoversion{bold}{\mathbfit}{cmr}{bx}{it}
  \newmathalphabet{\mathbfss} 
  \addtoversion{normal}{\mathbfss}{cmss}{bx}{n}
  \addtoversion{bold}{\mathbfss}{cmss}{bx}{n}
  \ifAMStwofonts
    \ifCUPmtlplainloaded \else
      \UseAMStwoboldmath
      \makeatletter
      \new@mathgroup\upmath@group
      \define@mathgroup\mv@normal\upmath@group{eur}{m}{n}
      \define@mathgroup\mv@bold\upmath@group{eur}{b}{n}
      \edef\UPM{\hexnumber\upmath@group}
      \new@mathgroup\amsa@group
      \define@mathgroup\mv@normal\amsa@group{msa}{m}{n}
      \define@mathgroup\mv@bold\amsa@group{msa}{m}{n}
      \edef\AMSa{\hexnumber\amsa@group}
      \makeatother
      \mathchardef\upi="0\UPM19
      \mathchardef\umu="0\UPM16
      \mathchardef\upartial="0\UPM40
      \mathchardef\leqslant="3\AMSa36
      \mathchardef\geqslant="3\AMSa3E

      \let\leq=\leqslant 
       \let\ge=\geqslant
    \fi
  \fi
\fi 

\ifnfsstwo
  \DeclareMathAlphabet{\mathbfit}{OT1}{cmr}{bx}{it}
  \SetMathAlphabet\mathbfit{bold}{OT1}{cmr}{bx}{it}
  \DeclareMathAlphabet{\mathbfss}{OT1}{cmss}{bx}{n}
  \SetMathAlphabet\mathbfss{bold}{OT1}{cmss}{bx}{n}
  \ifAMStwofonts
    \ifCUPmtlplainloaded \else
      \DeclareSymbolFont{UPM}{U}{eur}{m}{n}
      \SetSymbolFont{UPM}{bold}{U}{eur}{b}{n}
      \DeclareSymbolFont{AMSa}{U}{msa}{m}{n}
      \DeclareMathSymbol{\upi}{0}{UPM}{"19}
      \DeclareMathSymbol{\umu}{0}{UPM}{"16}
      \DeclareMathSymbol{\upartial}{0}{UPM}{"40}
      \DeclareMathSymbol{\leqslant}{3}{AMSa}{"36}
      \DeclareMathSymbol{\geqslant}{3}{AMSa}{"3E}

      \let\leq=\leqslant 
       \let\ge=\geqslant
    \fi
  \fi
\fi 

\ifCUPmtlplainloaded \else
  \ifAMStwofonts \else 
    \def\upi{\pi}
    \def\umu{\mu}
    \def\upartial{\partial}
  \fi
\fi

\title[The Optical QSO Luminosity Function]{The 2dF QSO Redshift Survey - I.  The Optical QSO Luminosity Function}

\author[B.J.Boyle et al.]
	{B.J.Boyle$^{1}$, T.Shanks$^{2}$, S.M.Croom$^{3}$, R.J.Smith$^{4}$,
	L.Miller$^{5}$, N.Loaring$^{5}$,
	C.Heymans$^{1}$\\  
${^1}$ Anglo-Australian Observatory, PO Box 296, Epping, NSW 1710, 
Australia \\ 
${^2}$ Department of Physics, University of Durham, South Road, 
Durham, DH1 3LE \\
${^3}$ Astrophysics Group, ICSTM, Blackett Laboratory, Prince Consort
Road, London, SW7 2BZ. \\
${^4}$ Research School of Astronomy and Astrophysics, ANU, Private
bag, Weston Creek P.O., ACT 2611, Australia.\\
${^5}$ Department of Physics, Oxford University, 1 Keble Road, Oxford,
OX1 3RH}

\begin{document}

\maketitle

\newcommand{\fmmm}[1]{\mbox{$#1$}}
\newcommand{\scnd}{\mbox{\fmmm{''}\hskip-0.3em .}}
\newcommand{\scnp}{\mbox{\fmmm{''}}}

\begin{abstract}

We present a determination of the optical QSO luminosity function and
its cosmological evolution with redshift for a sample of over 6000
QSOs identified primarily from the first observations of the 2dF QSO
Redshift Survey (2QZ).  For QSOs with $-26 < \mb < -23$ and $0.35 < z <
2.3$, we find that pure luminosity evolution (PLE) models provide an
acceptable fit to the observed redshift dependence of the luminosity
function.  The luminosity function is best fit by a two-power-law
function of the form
$\Phi(\lb)\propto[(\lb/\lb^*)^{\alpha}+(\lb/\lb^*)^{\beta}]^{-1}$.
Exponential luminosity evolution models, both as a function of look-back
time, $\lb ^{*}(z) = \lb^{*}(0){\rm e}^{k_{1}\tau}$, and as a general
second-order polynomial, $\lb^*(z)\propto 10\,^{k_1z+k_2z^2}$, were
found to provide acceptable fits to the dataset comprising the 2QZ 
and the Large Bright Quasar Survey.  Exponential
evolution with look-back time is prefered for $q_0=0.05$, while the
polynomial evolution model is prefered for $q_0=0.5$.  
The shape and evolution of the LF at low redshifts ($z<0.5$) and/or high
luminosities, not currently well sampled by the 2dF QSO survey, may show 
departures from pure luminosity evolution, but the results presented here
show that over a significant range of redshift, PLE 
is a good description of QSO evolution.
\end{abstract}
\section{Introduction}

The QSO optical luminosity function (OLF) and its evolution with 
redshift provides fundamental information on the overall demographics 
of the AGN population.  It provides constraints on the physical models 
for QSOs (Haehnelt \& Rees 1993; Terlevich \& Boyle 1993), information 
on models of structure formation in the early Universe (Efstathiou 
\& Rees 1988) and a picture of the ionizing UV/optical luminosity 
density from QSOs as a function of redshift (Meiksin \& Madau 1993; 
Boyle \& Terlevich 1998).

Based on the major ultra-violet excess (UVX) QSO surveys of the 1980s 
(Bracessi: Marshall et al.\ 1983, Palomar--Green: Green, Schmidt \&
Liebert 1986, 
Durham--AAT: Boyle et al.\ 1990), a picture emerged in which the 
low-intermediate redshift ($z < 2.2$) QSO OLF was modelled by a 
two-power-law function with a steep bright end ($\Phi(\lb)
\propto \lb ^{-3.6\pm0.1}$) and a much flatter faint end ($\Phi(\lb) 
\propto \lb ^{-1.2\pm0.1}$) whose redshift dependence was best fit by 
pure luminosity evolution, i.e., a uniform increase in luminosity 
toward high redshift (see e.g., 
Marshall 1985; Boyle et al.\ 1988; Hartwick \& Schade 1990). 
This evolution was modelled as a power-law in redshift of the functional 
form $\lb^*\propto (1 + z)^{k_{\rm L}}$, where $3 < k_{\rm L} < 3.4$ for an 
Einstein-de Sitter universe.

Later QSO surveys, particularly those that focussed on bright
magnitudes (LBQS: Hewett, Foltz \& Chaffee 1993, EQS: Goldschmidt \&
Miller 1998, HBQS: La Franca \& Cristiani 1997) have reported evidence
for a more complex form of evolution in which the bright end of the
OLF showed a significant steepening with increasing redshift.  Indeed
the steepening was sufficiently dramatic that, at the very lowest
redshifts studied, the OLF showed a single featureless power law
(Koehler et al.\ 1997) with little evidence for any cosmic evolution
amongst the brightest QSOs ($\mb < -27$).  Coupled with this, further
claims were made (Hawkins \& Veron 1995) that the observed break in
the luminosity function at higher redshifts was much less dramatic
than had previously been reported.

With the increasing number of high redshift ($z > 2$) QSOs discovered
in surveys with well-defined selection criteria, evidence was also
found that the strong power-law evolution does not continue on beyond
$z \sim 2$.  The nature of this change was modelled in a variety of
ways; from a simple halt to constant comoving density (see Boyle et
al.\ 1991) between $z \sim 2$ to $z \sim 3$ to a slackening off of the
evolution rate over the broad redshift range $1.6 < z < 3.3$ (Hewett
et al.\ 1993; Warren, Hewett \& Osmer 1994).  From surveys of higher
redshift QSOs ($3.5 < z < 5.0$), strong evidence emerged for the onset
of a dramatic decline in the QSO space density at $z > 3.5$ (see e.g
Warren et al.\ 1994).

However, compared to our knowledge of the galaxy luminosity function,
QSO OLF determinations were based on relatively few objects.  This was
particularly true for the extremes of the redshift distribution $z <
0.3$ and $z > 4$ or the high luminosity end of the OLF ($\mb < -27$),
where departures from the pure luminosity evolution model were first
noted.  In these regimes, the numbers of objects in suitable surveys
could be counted in the tens.  Even in the most well-sampled regions
of the QSO ($\mb, z$) plane, surveys comprised a few hundred or up to
one thousand QSOs in total.

The long-heralded arrival of the new generation of large QSO surveys
compiled with facilities such as the 2dF (Lewis et al. 1998) or
Sloan Digital Sky Survey (Loveday et al.\ 1998) will shortly generate
many tens of thousands of QSOs with which to carry out the definitive
study of the QSO OLF and its evolution with redshift.  These surveys
also provide the opportunity to study the QSO OLF as a function of
various physical parameters (e.g., emission-line strength, continuum
slope) or types (e.g., BAL QSOs).

In this paper, we present a new determination of the QSO OLF based on
the first $\sim 6000$ QSOs identified in the 2dF QSO redshift survey
(2QZ) (see Smith et al.\ 1998).  The 2QZ is currently more than a
factor of 10 larger than previous QSO surveys to a similar magnitude
limit ($\bj < 20.85$ mag).  When complete, it is planned that the 2QZ
will comprise 25000--30000 QSOs.  In section 2 we present a brief
overview of the survey and in section 3 we describe the analysis of
the survey.  We present our conclusions in section 4.

\section{data}

\subsection{The 2dF QSO redshift survey} 

\subsubsection{Input catalogue}

For the purposes of the analysis we have used the current version 
(as of September 1999) of the 2QZ catalogue containing 6684 QSOs in total. 
The identification of the QSO candidates for the 2QZ was based on 
broadband $u\bj r$ colour selection from APM measurements of UK 
Schmidt (UKST) photographic plates.  The survey area comprises 30 
UKST fields, arranged in two $75\degree \times 5\degree$ declination strips centred on 
$\delta = -30\degree$ and $\delta = 0\degree$.  The $\delta = -30\degree$ strip 
extends from $\alpha$ = 21$^{h}$40 to $\alpha$ = 3$^{h}$15 and the 
equatorial strip from $\alpha$ = 9$^{h}$50 to $\alpha$ = 14$^{h}$50. 
The total survey area is 740 deg$^{2}$, when allowance is made for 
regions of sky excised around bright stars.  The 2QZ area forms an 
exact subset of the 2dF Galaxy Redshift Survey (GRS: see Colless 1998)
area, with identical `holes' used for both surveys.  In a typical 
2dF field, approximately 225 fibres are devoted to galaxies, 125 to QSOs 
and 25--30 fibres are devoted to sky.  The data are reduced using the 
standard 2dF pipeline reduction system (Bailey \& Glazebrook 1999).

\begin{figure}
\vspace{-0.5cm}
\centering \centerline{\epsfxsize=9truecm
\epsfbox{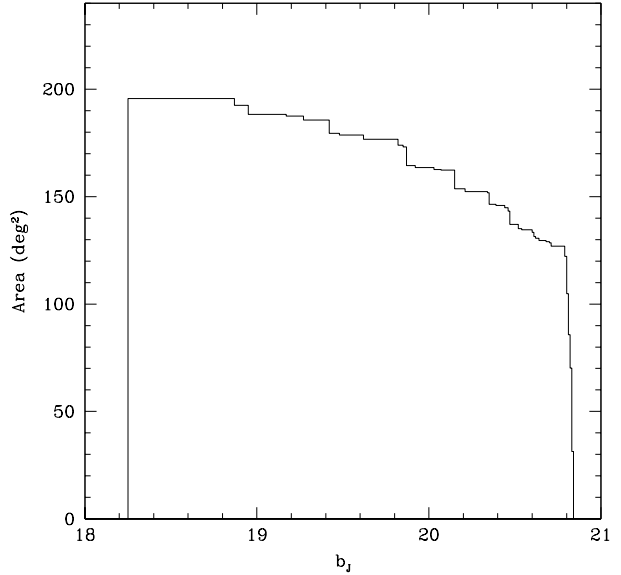}}\epsfverbosetrue
\caption{ Coverage function for the 2dF QSO redshift survey.}
\label{fig:coverage}
\end{figure}

\begin{figure}
\vspace{-0.5cm}
\centering \centerline{\epsfxsize=9truecm
\epsfbox{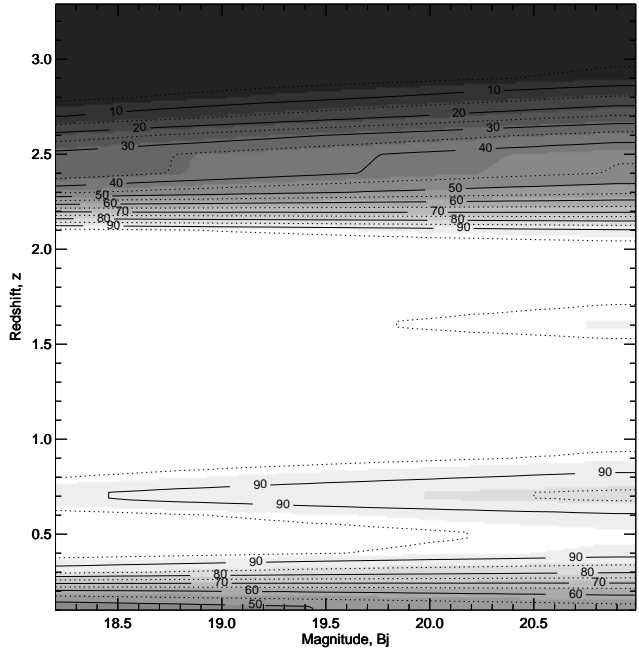}}\epsfverbosetrue
\caption{Photometric completeness contours in the ($\bj, z$) plane for 
the 2QZ.  The contours are plotted at 5 per cent intervals: 
dotted lines denote 5, 15, 25,...95 per cent completeness; solid 
lines correspond to 10, 20, 30...90 per cent.}
\label{fig:phcom}
\end{figure}

In each UKST field, APM measurements of one $\bj$ plate, one $r$ plate and 
up to four $u$ plates/films were used to generate a catalogue of 
stellar objects with $\bj < 20.85$.  A sophisticated procedure was 
devised to ensure catalogue homogeneity (Smith 1998).  Corrections were
made for vignetting and field effects due to variable desensitization
in the UKST plates, these effects being particularly noticeable at the
edges of plates.
The criteria for inclusion in the catalogue were ($u-\bj)\leq0.36; 
(u-\bj)<0.12-0.8(\bj-r); (\bj-r)<0.05$.  Based 
on the colours of QSOs which have previously been identified 
in the survey region, we estimate the catalogue is $\sim$ 90\% 
complete for $z < 2$ QSOs and comprises $\sim$55\% QSOs 
(see Croom 1997).  Subsequent spectroscopic observations with
2dF have confirmed this QSO fraction, with the principal contamination 
arising from galactic subdwarfs and compact blue galaxies.  Full details 
of the construction of the input catalogue may be found in Croom (1997) 
and Smith (1998).

\subsubsection{Spectroscopic observations}

\begin{figure*}
\vspace{-0.5cm}
\centering \centerline{\epsfxsize=12truecm
\epsfbox{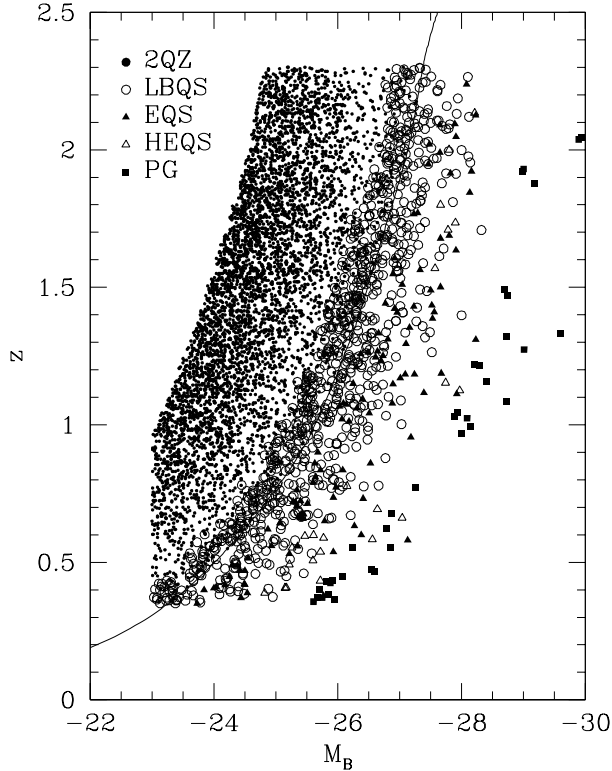}}\epsfverbosetrue
\caption{($\mb, z$) distribution for QSOs used in this analysis.
$\mb$ calculated using $q_0=0.5$,
$H_0=50$km\,s$^{-1}\,$Mpc$^{-1}$.  The symbols used for the different
surveys are indicated on the plot.  The bright limit for the 
2QZ survey, $\bj > 18.25$ mag, is denoted by the solid line.}
\label{fig:zmb}
\end{figure*}

2dF spectroscopic observations for the 2QZ began in January 1997, 
although the bulk of the redshifts have been obtained in the 
observing runs after October 1998 as the 2dF system has gained functionality 
(increased number of fibres, faster field re-configuration times).  Over
6000 QSO redshifts have been obtained with the 2dF, and the 2QZ 
is now the largest single homogeneous QSO catalogue in existence. 
Each 2dF field in the survey is observed for typically one hour. 
The spectral resolution is 4\AA~pixel$^{-1}$ and the spectra cover 
the wavelength range 3700\AA--8000\AA.  This set-up gives a typical 
signal-to-noise ratio of approximately seven or greater between 
4000\AA--6500\AA~in the continua of the faintest objects ($\bj = 20.85$) 
in the 2QZ.  This was sufficient to identify up to 85 per cent of the 
objects in each 2dF field down to this magnitude limit.  Where poor 
conditions prevented us from achieving this identification
rate in any single 2dF observation, we included in the final catalogue 
only those QSOs that were brighter than the magnitude at which 80 per 
cent of the QSO candidates in the field had a positive identification.

\subsubsection{Spectroscopic Incompleteness}
 
In the OLF analysis below, we correct for incompleteness in the 2QZ 
by using an effective area which is a function of both apparent $\bj$ 
magnitude and redshift.  Hereinafter, all $\bj$ magnitudes are corrected 
for galactic extinction using the Schlegel, Finkbeiner \& Davis (1998) values.

The 2QZ is subject to three forms of spectroscopic incompleteness.
First, a small fraction of survey fields were observed in poor
conditions, and do not reach the target spectroscopic completeness of
80 per cent.  As described above, the magnitude limits for these
fields were made brighter until the desired completeness level was
reached.  Secondly, we corrected the inevitable trend to increasing
spectroscopic incompleteness at fainter magnitudes by correcting the
actual area observed by the fraction of objects with a reliable
spectroscopic identification.  This was done in 0.025mag bins to track
accurately the magnitude-dependent incompleteness.

Finally, we applied a uniform correction factor of 0.71 to the
effective  area to account for the 29 per cent objects in the input
catalogue that  have not yet had been observed by 2dF, despite being
located in areas  already observed in the 2dF survey.  This is because
the numbers of candidates  in the combined QSO/galaxy catalogues in a
typical 2dF field is $\sim$ 50  per cent higher than the number of
fibres available.  Near-complete ($>$ 95  per cent) spectroscopic
coverage of the survey area by 2dF is achieved  by a complex tiling
algorithm, with each 2dF field being visited 1.5  times on average.
Since the observations are still at a relatively  early stage, the
coverage of the survey area with 2dF is relatively patchy and
the completeness fraction is still relatively low (71 per cent).  The
resulting effective area of the survey is plotted in
Figure~\ref{fig:coverage}.  

\subsubsection{Photometric Incompleteness}
 
Photometric incompleteness, arising from the errors in
the photographic magnitudes ($\pm$0.1 mag in each band) and
variability (due to the noncontemparaneous nature of the $u\bj r$
plates on each field) will cause QSOs to exhibit $u-\bj/\bj-r$ colours
outside our selection criteria.  The incompleteness is a complex
function of both magnitude and redshift.  An estimate of this
incompleteness has been made by Croom et al.\ (in preparation) using mean
colour-redshift relations (zero-pointed to the 2QZ system) from the
non-colour-selected QSO survey of Hawkins \& Veron (1995).  At
$z<2.3$ the Hawkins \& Veron colours accurately trace the mean colours
of the 2QZ QSOs.  The dispersion in the colours as a function of
magnitude was derived from the observed dispersion of colours for the
2QZ QSOs in the redshift interval $1 < z < 2$, the regime of highest
completeness.  A Monte Carlo simulation using the measured dispersion
in colours and the Hawkins \& Veron mean colours, with $10^{6}$ QSOs
in each $\Delta\bj$ and $\Delta z$ bin, was then used to predict the
photometric completeness as a function of both magnitude and redshift.
The completeness contours are shown in Figure~\ref{fig:phcom}.  The
photometric completeness is largely independent of magnitude and is at
least 85 per cent or greater over the redshift range $0.4 < z <
2.1$.  At higher redshifts the completeness rapidly drops, falling to
below 50 per cent at $z > 2.3$.  We have therefore chosen this
redshift as the upper limit to our analysis below.  Although the
catalogue contains many hundreds of QSOs with $z > 2.3$ (which will be
used for the clustering analysis that is not so dependent on
photometric completeness), any small errors in the completeness
estimates at these redshifts can lead to large variations in the
computed effective area.

At the lowest redshifts, the derived photometric completeness is still
relatively high ($\sim$ 60 per cent at $z = 0.2$).  However, low
redshift, low luminosity QSOs ($\mb>-23$) dominated by their host galaxy
light will be lost from the 2QZ as a result of the stellar selection
criterion applied to the input catalogue.  Although it may be possible to
correct for such selection effects, for the present analysis we have
simply chosen to exclude all low luminosity QSOs ($\mb>-23$) from the
OLF fitting procedure below.  Given the bright apparent magnitude
limit of the 2QZ ($\bj=18.25$ mag), this effectively limits the
minimum redshift in the complete 2QZ sample to $z>0.35$.  We have
therefore adopted this as our low redshift limit for all QSO surveys
used in the OLF analysis below.  Thus, out of a total of 6684 QSOs
identified in the 2QZ (13552 objects observed in 219 2dF fields), 5067
fulfil the criteria for inclusion in the complete sample ($q_0=0.5$,
$H_0=50$km\,s$^{-1}\,$Mpc$^{-1}$).

\begin{table*}
\centering
\caption{Parameters for the surveys used in our analysis.  $N_{\rm
 QSO}$ is the number of QSOs with $\mb<-23$ ($q_0=0.5$, 
$H_0=50$km\,s$^{-1}$\,Mpc$^{-1}$) in the redshift interval $0.35<z<2.3$.}
\label{tab:surv}
\begin{tabular}{@{}lrcrl@{}}
\hline
Survey&$N_{\rm QSO}$&$B_{\rm lim}$&Area&Reference\\
&&&(deg$^2$)\\
\hline
2dF & 5057 & $18.25 < \bj < 20.85$ & 196 & This paper\\
LBQS & 867 & $16.50 < \bj < 18.85$ & 454 & Hewett et al.\ (1995)\\
EQS  & 106 & $15.00 < \bj < 18.00$ & 330 & Miller et al.\ (unpublished)\\
HEQS & 23  & $15.00 < \bj < 17.65$ & 611 & Koehler et al.\ (1997)\\ 
PG & 36 & $13.00 < \bj < 16.67$ & 10653 & Green et al.\ (1986)\\ 
\hline
\end{tabular}
\end{table*}

\subsection{Other Samples}

We are currently in the process of extending the bright limit of the
2QZ from $\bj=18.25$ mag to $\bj=17$ mag using observations made with
the FLAIR spectrograph on the UK Schmidt Telescope.  These
observations are not yet complete and we have incorporated a number of
independent QSO surveys at brighter magnitudes to extend the present
analysis of the OLF to higher luminosities.  Table~\ref{tab:surv}
lists the areas, magnitude limits, and the number of QSOs within the
completeness limits ($\mb<-23$, $0.35<z<2.3$) for each QSO survey used
in this paper. The absolute magnitudes and redshifts for the QSOs
in these surveys are plotted in Figure~\ref{fig:zmb}.

The Large Bright QSO Survey (LBQS: Hewett et al.\ 1995) provides a
complementary sample to the 2QZ.  The details and completeness of the
survey are well established and it provides a large number of QSOs
in the $1.5 - 2\,$mag interval brighter than the $\bj>18.25$ 2QZ
survey limit.  

At the brightest magnitudes $B < 16.5\,$mag, the largest currently
published survey is the Palomar--Green survey (Green et al.\ 1986).
However, the completeness of this survey has been called into question
by a number of recent surveys including the Edinburgh Quasar Survey
(EQS: Goldschmidt et al.\ 1998) and the Hamburg/ESO Quasar Survey
(HEQS: Koehler et al.\ 1997).  Unfortunately, full details of both these
catalogues have yet to be published, and although we have access to
the unpublished EQS, details of its completeness have yet to be
accurately determined.

Given the remaining uncertainty over their details, and apparent
discrepancy between QSO surface densities derived from
these brighter surveys, we chose to carry out our
analysis of the OLF for three separate survey combinations: a) 2QZ +
LBQS; b) 2QZ + LBQS + PG; c) 2QZ + LBQS + EQS + HEQS.  Given the
available information, the 2QZ + LBQS data sample constitutes our
primary data sample in this paper.  When more extensive published
catalogues are available from brighter surveys (in particular the
HEQS) they will clearly provide a powerful test of the OLF models at the
brightest absolute magnitudes (see Figure~\ref{fig:zmb}).

A further survey at intermediate magnitudes, the Homogeneous Bright
QSO Survey (HBQS) is currently under construction (see Cristiani et
al.\ 1995).  There is considerable overlap between the survey areas
for the HBQS and EQS.  The currently published HBQS samples contains
fewer QSOs than the EQS sample to which we have access.  We have
therefore chosen not to include the HBQS in the analysis below.

With one exception, the surveys are all largely independent of one another.
The the current coverage of the 2QZ results in a very small overlap with the 
LBQS; there is only one QSO in common between the 2QZ and LBQS.  Given the 
complexity of the current 2dF coverage function in the spatial domain 
(due to incomplete tiling), we therefore chose to treat the LBQS and 2QZ 
as independant surveys. At brighter magnitudes, the PG, HEQS and EQS are 
all independant of one another.  

The only exception is the overlap between the LBQS and EQS.  There are
32 QSOs in the EQS sample which are also in the LBQS sample, or
approximately 30 per cent of the EQS sample used in this analysis.
The QSOs are distrubuted throughout the four UK Schmidt fields in
common between the two surveys.  In the analysis below we have
therefore excluded these four fields from the EQS during the model
fitting procedure, but we test the acceptibilty of the best-fit model
against the full EQS sample. The number of QSOs for the EQS quoted in Table 1
refers to the full sample.
  
\begin{figure*}
\centering \centerline{\epsfxsize=19truecm
\epsfbox{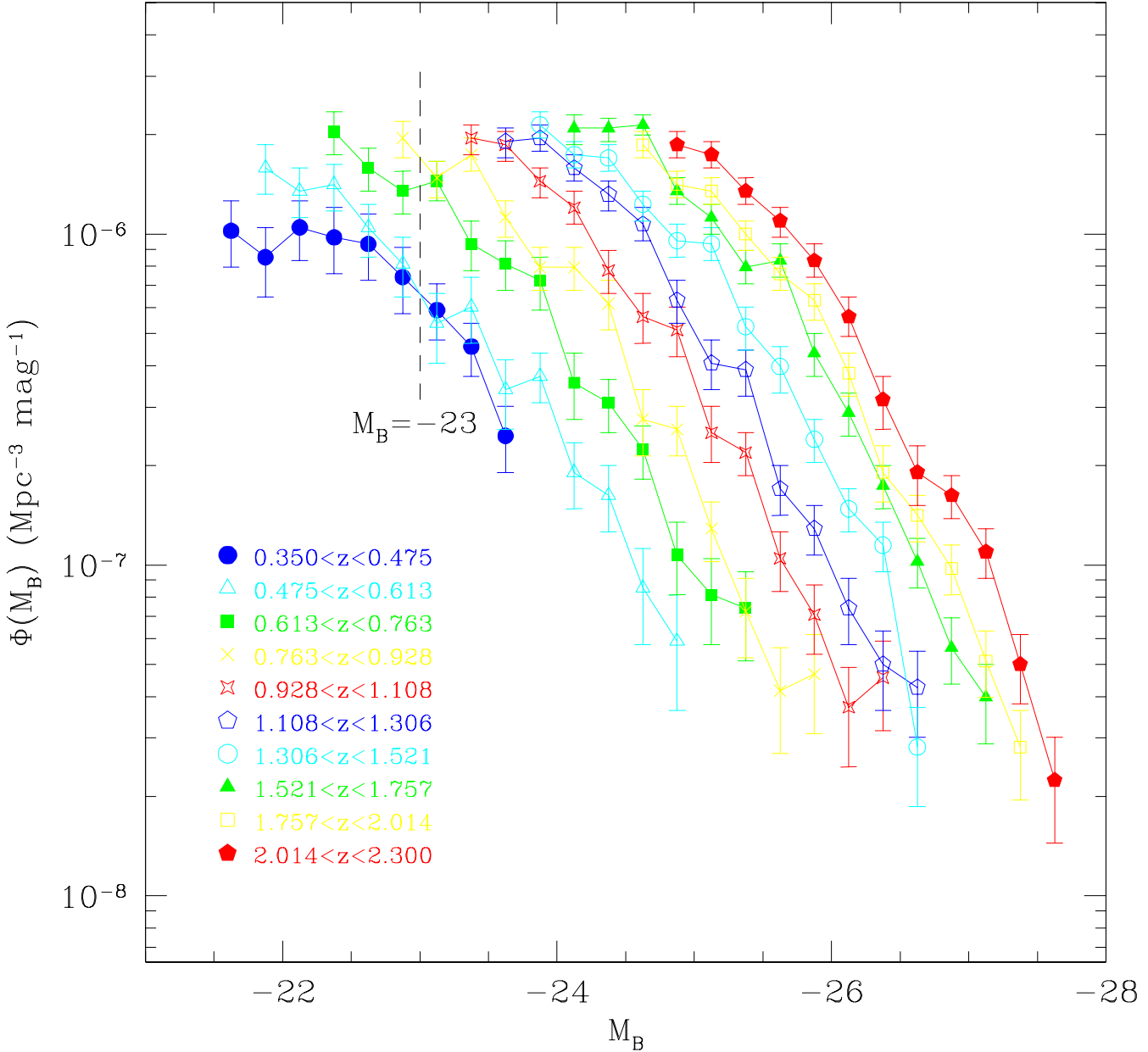}}\epsfverbosetrue
\vspace{-1.0truecm}
\caption{Luminosity function for the 2QZ + LBQS dataset in a flat
$q_0=0.5$ universe}
\label{fig:olf} 
\end{figure*}

\section{ANALYSIS}

\subsection{$1/V$ estimator}

As the first step in our analysis, we obtained a graphical
representation of the OLF and its evolution with redshift. We used the
new $1/V$ estimator devised by Page \& Carrera (1999) to derive a
binned estimate of the OLF. 
Although it addresses one of the potential biases of the traditional
$1/V_a$ statistic (Schmidt 1968), like most binned LF estimators the
Page \& Carrera (1999) method assumes that absolute magnitude and
redshift bins can be chosen sufficently small so that the effects of
evolution and a steeply rising OLF are negligable across each bin. For
previous samples this has been difficult to achieve whilst still
retaining a sufficient number of QSOs in each bin.

With the large numbers of QSOs now available for this analysis,
we were able to minimise these effects by choosing much
smaller bins in redshift and absolute magnitude.
We used 0.25 mag bins and 10 redshift bins equally 
spaced in $\log (1+z)$ over the interval $0.35 < z < 2.3$
to compute the $1/V$ estimate of the OLF.
  

For an OLF that evolves significantly across a $M_B$,$z$ bin, $1/V$
methods will still yield biased estimates of the OLF in bins cut by
the magnitude limit of the sample. In such bins, the estimated OLF
will only represent the mean space density of objects sampled in
bright $M_B$, low $z$ region of the bin.  In the absence of any
magnitude limit most objects would tend to lie in the faint $M_B$,
high $z$ part of the bin (at least for a power law LF undergoing
strong redshift evolution) and so the estimate of the OLF in bins
which are not fully sampled will be
biased low.  The extent of the bias will, of course, depend on the
extent to which the OLF evolves across the bin.  Our choice of small
bins alleviates, but does not remove this bias and so we have also
excluded all $M_B,z$ bins which contain the $\bj\sim 20.8$ mag survey
limit in our binned estimate of the OLF.  A similar bias also affects
(but in the opposite sense) bins cut by the bright mangitude limit of the 
sample. However, we also imposed the constraint that we would not 
compute the OLF in bins where there were five or fewer objects.  
This latter conditions prevents any bins cut by the bright survey
limit from appearing in estimate of the OLF below.   

The resulting OLF, calculated for a flat universe with $q_0=0.5$ and a
$H_{0} = 50 {\rm km\,s}^{-1}\,{\rm Mpc}^{-1}$, in plotted
Figure~\ref{fig:olf}.  Absolute $\mb$ magnitudes were derived for the
QSOs using the k-corrections derived by Cristiani \& Vio (1990).  The
median number of QSOs in each plotted bin is 40, although some bins
contain up to 170 QSOs.  The shape and redshift dependence of the OLF
in Figure~\ref{fig:olf} are strongly suggestive of a luminosity
evolution model similar to those derived previously from the
Durham/AAT sample (Boyle et al.\ 1988).


\subsection{Maximum Likelihood Analysis}

\subsubsection{Method}

To obtain a more quantitative descriptions of suitable models, we
carried out a maximum likelihood fitting procedure for a number of
models to the data (see Boyle et al.\ 1988). This technique relies on
minimizing the likelihood function $S$ corresponding to the Poisson
probability distribution function for both model and data (Marshall et
al.\ 1983).  We tested the goodness-of-fit of the model to the data
using the 2D Kolmogorov-Smirnoff (KS) statistic.  The KS is
notoriously insensitive to discrepancies between the data and the
model predictions in the wings of the distributions.  This problem is
particularly severe in the present analysis, where the weakness of the
KS test at the brightest absolute magnitudes is compounded by the
vanishingly small fraction of objects in this region of the $M_B,z$
plane in the combined sample (see Figure~\ref{fig:zmb}).

To alleviate this problem, we derived 2D KS statistics from each
survey separately; combining them into a final KS probability using
the $Z$ statisitic described by Peacock (1983).  This approach enables
each of the model predictions to be tested against the data in specific
regions of the ($M_{\rm B}, z$) plane sampled by different surveys.  
A significant rejection of the model, even by
a relatively small sample, will thus have major impact on the overall
acceptability of the fit.

In the fitting procedure, we used no more free parameters in any model
than was required to obtain an acceptable fit. We defined $a\ priori$
an acceptable fit as one which could not be rejected at the 99 per
cent confidence level or greater i.e., a KS probability ($P_{\rm KS}$)
of 1 per cent or greater.  Errors on the fit parameters correspond to
the $\Delta S = 1$ contours around each parameter, or, equivalently,
the 68 per cent confidence contour for one interesting parameter.

\begin{table*}
\centering \begin{minipage}{200mm} \caption{Best-fit OLF model parameters from maximum likelihood analysis.}
\label{tab:params}

\begin{tabular}{@{}cccccccccccc@{}}
\hline
 Surveys sampled & Evolution & $q_{0}$ &  $N_{\rm QSO}$ &
$\alpha$ & $\beta$ & $M ^{*}_{\rm B}$ & $k_{1}$ & $k_{2}$ &  $\Phi^{*}$ &
$P_{\rm KS}$ \\

& Model & & & & & & & &  ${\rm Mpc}^{-3}{\rm mag}^{-1}$ &\\

\hline 
2dF \& LBQS & $exp(k \tau)$ &   0.05 &     6100 & 3.37 & 1.55 & --21.16 &7.11 & --- &   0.47$\times 10^{-6}$ &   0.52$\times 10^{-1}$\\ 
\"{}        & $exp(k \tau)$ &   0.50 &     5924 & 3.32 & 1.39 & --19.67 &6.98 & --- &   0.15$\times 10^{-5}$ &   0.14$\times 10^{-3}$\\
\"{}        & $k_{1}z + k_{2}z^{2}$&0.05 & 6100 & 3.29 & 1.47 & --22.49 &1.30 & --0.25 &0.56$\times 10^{-6}$ &   0.24$\times 10^{-2}$ \\ 
\"{}        & $k_{1}z + k_{2}z^{2}$&0.50 & 5924 & 3.43 & 1.58 & --21.99 &1.35 & --0.27 &0.11$\times 10^{-5}$ &   0.12\\ 

2dF, LBQS \& PG& $exp(k \tau)$ &0.05 &     6136 & 3.37 & 1.63 & --21.33 & 6.98 & ---&    0.41$\times10^{-6}$ &   0.14$\times 10^{-1} $ \\
\"{} & $exp(k \tau)$        &   0.50 &     5960 & 3.52 & 1.57 & --20.00 & 6.95 & ---&    0.10$\times10^{-5}$ &   0.32$\times 10^{-6}$ \\ 
\"{} & $k_{1}z + k_{2}z^{2}$   &0.05 &     6136 & 3.30 & 1.55 & --22.33 & 1.42 &--0.29 & 0.52$\times10^{-6}$ &   0.34$\times 10^{-2}$ \\ 
\"{} & $k_{1}z + k_{2}z^{2}$   &0.50 &     5960 & 3.60 & 1.77 & --22.39 & 1.31 &--0.25 &0.68$\times 10^{-6}$ &   0.15$\times 10^{-1}$ \\ 

2dF,LBQS, HEQS, \& EQS&$exp(k \tau)$&0.05 &6265 & 3.95 & 1.87 & --21.80 & 7.16 & ---&   0.17$\times 10^{-6}$ &   0.28$\times 10^{-1}$ \\ 
\"{} & $exp(k \tau)$            &   0.50 & 6089 & 3.52 & 1.57 & --20.00 & 6.95 & ---&   0.10$\times 10^{-5}$ &   0.10$\times 10^{-5}$ \\ 
\"{} & $k_{1}z + k_{2}z^{2}$   &    0.05 & 6265 & 3.39 & 1.47 & --22.55 & 1.30 &--0.25 &0.52$\times 10^{-6}$ &   0.60$\times 10^{-4}$ \\
\"{} & $k_{1}z + k_{2}z^{2}$   &   0.50 &  6089 & 3.49 & 1.53 & --21.98 & 1.34 &--0.27 &0.61$\times 10^{-6}$ &   0.83$\times 10^{-3}$ \\
\hline

\end{tabular}
\end{minipage}
\end{table*} 

\subsubsection{Models}

Guided by the appearance of the OLF in Fig~\ref{fig:olf} we chose to
model the OLF $\Phi(L, z)$ as a two-power-law in
luminosity,\footnote{All simpler forms of the OLF e.g., single power
law, Schechter function were strongly ruled out ($P_{\rm KS} <
10^{-10}$) in preliminary fits.}

\begin{displaymath}
\Phi (\lb ,z) = \frac{\Phi(\lb ^{*})}{[(\lb /L_{\rm B}^{*})^{\alpha} 
+ (\lb /\lb ^{*})^{\beta}]}.
\end{displaymath}
Expressed in magnitudes this becomes
\begin{displaymath}
\Phi (\mb,z) = \frac{\Phi(\mb^{*})}{10^{0.4[(\alpha+1)(\mb-
\mb^{*}(z))]} + 10^{0.4[(\beta+1)(\mb-\mb^{*}(z))]}},
\end{displaymath}
where the evolution is given by the redshift dependence of the break
luminosity $\lb^*$ or magnitude, $\mb^{*}(z)$.  

We first attempted to fit the evolution by using `standard' models,
i.e., an exponential luminosity evolution with fractional look-back
time ($\tau$), such that:
\begin{displaymath}
\lb ^{*}(z) = \lb ^{*}(0)\exp(k_{1}\tau),
\end{displaymath}
and a power-law luminosity evolution model with a redshift cut-off:
\begin{displaymath}
\lb ^{*}(z) = \lb ^{*}(0)(1 + z)^{k_{1}} \hspace{2cm}z< z_{\rm max}
\end{displaymath}
\begin{displaymath}
\lb ^{*}(z) = \lb ^{*}(0)(1 + z_{\rm max})^{k_{1}} \hspace{1.5cm} z\ge
z_{\rm max}
\end{displaymath}

In contrast to previous analyses, the power-law evolution model
provided a poor fit to all datasets ($P_{\rm KS}\ll0.01$) for all
cosmological models.  This implied that any decline in the evolution
at high redshifts $(z\sim 2$) was not well represented by an abrupt
cut-off at $z_{\rm max}$.  We therefore adopted a more general
exponential evolution law incorporating a second-order polynomial
function (hereinafter referred to as polynomial evolution) 
that gives a smoother transition in the power law behaviour
of the evolution at $z > 2$ as allows for the possibility of negative
evolution at high redshift \footnote{Polynomial evolution models were
first fit to the radio galaxy/QSO LFs by Dunlop \& Peacock (1990)}:
\begin{displaymath}
\lb ^{*}(z) = \lb ^{*}(0)10^{k_1z+k_2z^2},
\end{displaymath}
or equivalently,
\begin{displaymath}
\mb^{*}(z) = \mb^{*}(0) - 2.5 (k_1 z + k_2z^2).
\end{displaymath}
\subsubsection{Results}

The best-fit parameters, and KS probabilities for the polynomial and
exponential evolution fits are given in Table~\ref{tab:params}.  The
statistical errors on individual parameters are
$\Delta\alpha \sim \pm 0.05$, $\Delta\beta \sim \pm 0.1$, $\Delta
\mb^* \sim \pm 0.2$, $\Delta k_1 ({\rm exponential}) \sim \pm 0.1$,
$\Delta k_1 ({\rm polynomial}) \sim\pm0.05$, $\Delta k_2 \sim \pm 0.02$.

Acceptable fits ($P_{\rm KS} > 0.01$) to the primary 2QZ + LBQS sample
were found for both evolution models.  The exponential model favoured
low $q_{0}$ and the polynomial model favoured high $q_{0}$.  

\begin{figure*}
\centering \centerline{\epsfxsize=15truecm
\epsfbox{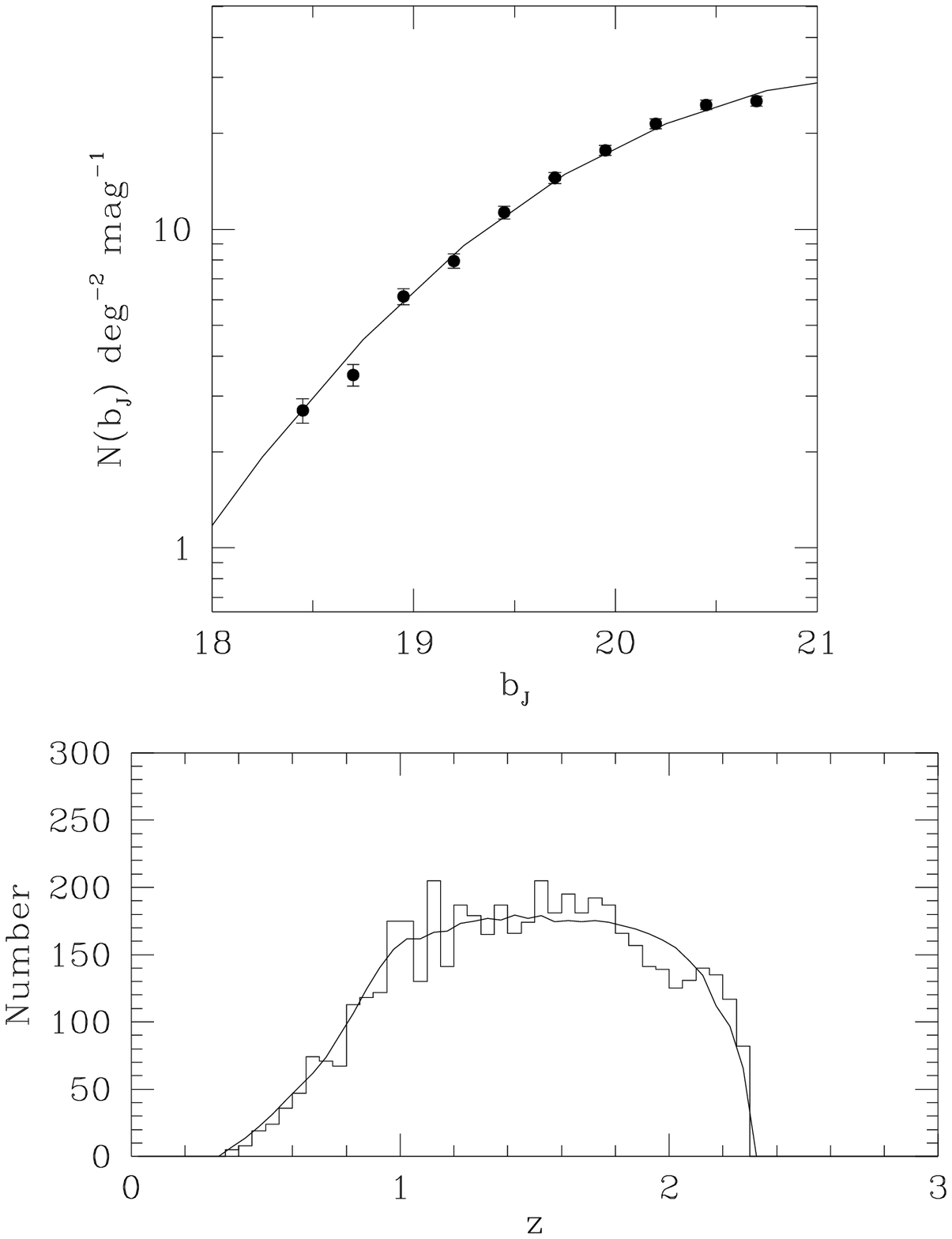}}\epsfverbosetrue
\vspace{-0.5truecm}
\caption{Upper panel: Derived differential number-magnitude, n(m),
relation for the 2QZ survey (filled dots) and prediction from best-fit
polynomial evolution model ($q_0=0.5$) to LBQS+2QZ sample.  Lower
panel: Observed and predicted number-redshift, n(z), relation for the
2QZ survey.  Model as for upper panel.  Note that the derived n(m)
relation from the 2QZ survey has been corrected for both spectroscopic
and photometric completeness, whereas the n(z) relation corresponds
simply to the observed numbers of QSOs in the 2QZ survey.
Correspondingly the predicted n(m) relation comes directly from the
model, whereas the n(z) model prediction has been corrected for the 
assumed survey incompleteness.}
\label{fig:nmnz} 
\end{figure*}

The predicted differential number-magnitude, n(m), and
number-redshift, n(z), relations for the 2QZ survey based on the
best-fit $q_0=0.5$ polynomial evolution model to the 2QZ + LBQS
samples are shown in Figure~\ref{fig:nmnz}.  The model predictions
provide a good fit to the derived n(m) and observed n(z)
relations for QSOs with $0.35<z<2.3$ and $M_B<-23$ from the 2QZ survey
also plotted in this figure.  

\begin{figure}
\centering \centerline{\epsfxsize=9truecm
\epsfbox{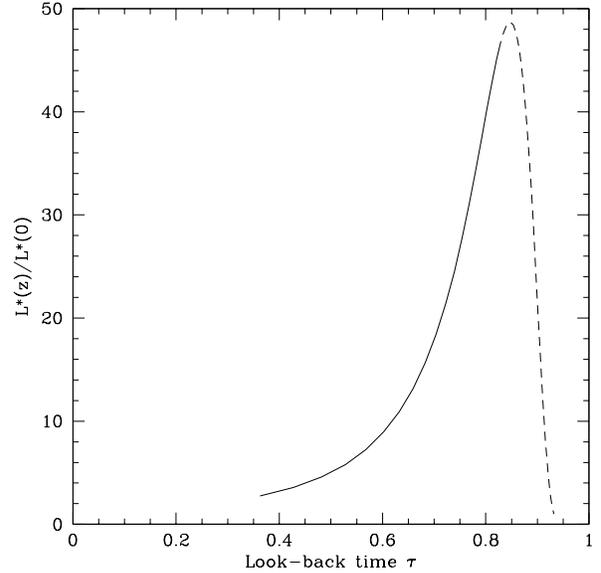}}\epsfverbosetrue
\vspace{-0.5truecm}
\caption{Evolution of $L_B^*$ as a function of 
fractional look-back time ($\tau$) for the polynomial evolution model
in a $q_0=0.5$ universe.  The solid line denotes the best-fit evolution
model obtained from fitting to the 2QZ + LBQS data-sets.  The
dashed line represents the extrapolated evolution of $L_B^*$ 
between $z=2.3$ and $z=5$.}
\label{fig:lstar} 
\end{figure}

The extrapolated maximum $\lb ^*$ in the polynomial evolution model occurs
in the range $2.46<z<2.50$ for $0.05<q_0<0.5$. The behaviour of $\lb ^*$
as a function of fractional look-back time for the $q_0=0.5$ universe 
is shown in Figure~\ref{fig:lstar}.  

\begin{figure}
\centering \centerline{\epsfxsize=9truecm
\epsfbox{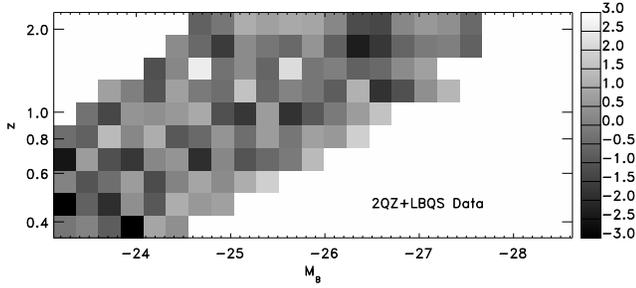}}\epsfverbosetrue
\vspace{.0truecm}
\caption{Greyscale respresentation of the Poisson significance 
of the discrepancy ($(N_{\rm obs}-N_{\rm pred})/\sqrt{N_{\rm pred}}$) between 
the observed number of QSOs ($N_{\rm obs}$) in the 2QZ + LBQS
data-set and predicted number of QSOs ($N_{\rm pred}$) 
from $q_0=0.5$ polynomial evolution model. The $M_B,z$ bins chosen
are the same as those used in the $1/V$ determination of the OLF. 
Only bins with $N_{\rm pred}>5$ have been plotted.}
\label{fig:resid} 
\end{figure}

In Figure~\ref{fig:resid} we have also plotted the Poisson
significance of the residual difference between the best-fit $q_0=0.5$
polynomial evolution model and 2QZ + LBQS data-set for the $M_B,z$
bins used in the $1/V$ analysis of the OLF.  It can be seen that,
within the range of luminosity-redshift parameter space covered by the
data, there are no significant ($>3\sigma$) differences between the
model and data.  Although there may be a possible weak trend at low
redshifts for the model to over-predict the numbers of QSOs at low
luminosities and to under-predict numbers at high luminosities, any
such discrepancies are not yet statistically significant with this
current dataset.

We found no significant difference in the values of the best-fit
parameters when the fits were restricted to $z>0.5$ or $z<2.0$,
i.e., excluding the sample regions where we apply the largest
correction for incompleteness.  The extrapolated peak in $\lb ^*$ occurs at the
same redshift ($z=2.5$) for model fits to both the $z<2$ and $z<2.3$
samples see (Figure~\ref{fig:lstar}).

The addition of the bright samples yielded much poorer overall
fits.  Inclusion of the PG sample gave, at best, marginally
acceptable fits for the exponential evolution model ($q_0=0.05$) and
polynomial evolution model ($q_0=0.5$).  No acceptable fit was
found for a $q_0=0.5$ universe with the incorportion of
HEQS and EQS, and only a barely acceptable fit ($P_{\rm KS} = 0.01$)
was obtained for the  $q_0=0.05$ exponential evolution model.

Previous attempts to characterize departures from pure luminosity evolution
(PLE) have led to models with a redshift-dependent bright-end slope
(La Franca \& Cristiani 1997).  We attempted to fit such a model
for the polynomial evolution models to the data-set including the
HEQS and EQS samples using a simple redshift dependence of the form
\begin{displaymath}
\alpha(z) = \alpha(0) + \kappa_3 z.
\end{displaymath} 
For $q_0=0.5$ the best fit value of $\kappa_3=0.36$, while for $q_0=0.05$,
$\kappa_3=0.0$.  However, this model did not provide an acceptable fit
to the data-set ($P_{\rm KS}<0.01$) for either $q_0=0.5$ or $q_0=0.05$.

By virtue of its size, rejection of the fits by the brighter data-sets
was dominated by the EQS. Nevertheless the EQS contains relatively few
QSOs, and a full understanding of the evolution of the very brightest
QSOs clearly still requires both larger samples and a more detailed
knowledge of their properties.  This should be available with the
completed HEQS.

There could be a variety of reasons why luminous QSOs may exhibit
deparatures from pure luminosity evolution models.  For example,
bright QSO samples are known to contain a greater fraction of
radio-loud QSOs (Peacock, Miller \& Longair 1986) and also contain a
greater fraction of gravitationally lensed QSOs (Kochanek 1991). Both
effects could give rise to systematic departures from simple
luminosity evolution models fit predominantly to less luminous QSOs.
Equally, these results may imply that luminous QSOs simply evolve
differently from the bulk of 'normal' QSOs.


\begin{table*}
 \centering
 \begin{minipage}{120mm}
  \caption{Parameter values for the polynomial evolution model as a function of $\Omega_{\rm m}$ and 
$\Omega_{\Lambda}$}
\label{tab:coscon}
\begin{tabular}{@{}ccccccccccccc@{}}
\hline
$\Omega_{\rm m}$ & $\Omega_{\Lambda}$ &  $N_{\rm QSO}$ & $\alpha$ & $\beta$ &
$M^{*}_{\rm B}$ & $k_{1}$ & $k_{2}$ & $\Phi^{*}$ & $P_{\rm KS}$ \\
 & & & & & & & & ${\rm Mpc}^{-3}{\rm mag}^{-1}$ \\
\hline
0.3 & 0.7 & 6180 & 3.41 & 1.58 & --22.65 & 1.36 &   --0.27 & 0.36$\times 10^{-6}$ &   0.075\\
0.5 & 0.5 & 6102 & 3.56 & 1.67 & --22.56 & 1.35 &   --0.27 & 0.44$\times 10^{-6}$ &   0.10\\
0.7 & 0.3 & 6042 & 3.52 & 1.66 & --22.38 & 1.34 &   --0.26 & 0.51$\times 10^{-6}$ &   0.098\\
0.1 & 0.0 & 6104 & 3.43 & 1.69 & --22.71 & 1.38 &   --0.28 & 0.35$\times 10^{-6}$ &   0.021\\
0.2 & 0.0 & 6094 & 3.56 & 1.72 & --22.74 & 1.35 &   --0.27 & 0.36$\times 10^{-6}$ &   0.032\\
0.3 & 0.0 & 6078 & 3.43 & 1.62 & --22.46 & 1.35 &   --0.27 & 0.54$\times 10^{-6}$ &   0.061\\
0.4 & 0.0 & 6060 & 3.65 & 1.80 & --22.74 & 1.35 &   --0.26 & 0.36$\times 10^{-6}$ &   0.093\\
0.5 & 0.0 & 6038 & 3.53 & 1.69 & --22.47 & 1.34 &   --0.26 & 0.56$\times 10^{-6}$ &   0.088\\
0.6 & 0.0 & 6017 & 3.31 & 1.49 & --22.05 & 1.33 &   --0.26 & 0.10$\times 10^{-5}$ &   0.095\\
0.7 & 0.0 & 5993 & 3.40 & 1.54 & --22.12 & 1.33 &   --0.26 & 0.94$\times 10^{-6}$ &   0.10 \\
0.8 & 0.0 & 5978 & 3.53 & 1.67 & --22.28 & 1.34 &   --0.26 & 0.76$\times 10^{-6}$ &   0.087\\
0.9 & 0.0 & 5946 & 3.42 & 1.56 & --22.05 & 1.32 &   --0.26 & 0.11$\times 10^{-5}$ &   0.066\\
1.0 & 0.0 & 5924 & 3.45 & 1.63 & --22.10 & 1.31 &   --0.26 & 0.10$\times 10^{-5}$ &   0.049\\
\hline
\end{tabular}
\end{minipage}
\end{table*} 

We also explored the effect of different cosmological models on the
OLF for our primary 2QZ + LBQS sample.  The results are reported in
Table~\ref{tab:coscon}.  For an $\Omega_{\Lambda} = 0$ universe we
fit the polynomial evolution model to the data for a variety
of different values of $q_0$.  All values of $q_0$ in the range
$0<q_0<0.5$ yielded acceptable fits, with probabilty reaching
a broad maximum in the the region $q_0=0.3-0.4$. However, this 
value is highly dependent on the evolution model chosen to fit the 
data.  For example the exponential evolution model favours lower 
values of $q_0$ ($q_0<0.1$, see Table 2). It is unlikely that the 
QSO OLF can be used to define useful contraints on the value of 
$q_0$ until a meaningful physical model for QSO evolution is available.

The best-fit OLF parameters for a non-zero cosmological 
constant in a flat universe (i.e., $\Omega_{\rm
m}+\Omega_\Lambda=1$) were also obtained using the expressions
for comoving distance $r(z)$ and comoving volume element $dV/dz$ 
given by:
\begin{displaymath}
r(z) = \frac{c}{H_{0}a_{0}} \int^{z}_{0} 
\frac{dz}{ \sqrt { \Omega_{\rm m}(1+z)^{3}+\Omega_\Lambda}}
\end{displaymath}
\begin{displaymath}
\frac{dV}{dz}(z) = r(z)^{2}\sqrt
{ \Omega_{\rm m}(1+z)^{3}+\Omega_{\Lambda}}
\end{displaymath}
A variety of fits for different values of $\Omega_M$ and $\Omega_\Lambda$ 
are presented in Table~\ref{tab:coscon}.  Luminosity evolution 
remains a good fit to the 2QZ + LBQS dataset in all cases.
As above, it is apparent the OLF provides little potential for 
discrimination between cosmological models with a non-zero 
$\Omega_\Lambda$.\\

\section{CONCLUSIONS}

For absolute magnitudes $-26<M_B<-23$ ($q_0=0.5$) and redshifts $0.35<z<2.3$,
pure luminosity evolution (PLE) has been shown to produce an
acceptable fit to the QSO distribution.  This region of parameter
space contains the QSOs reponsible for the vast majority of the AGN
luminosity density in the redshift range between $0.35<z<2.3$.  The
best-fitting PLE models exhibit an exponential increase in luminosity
with either look-back time ($q_0=0.05$) or as a second-order
polynomial function of redshift ($q_0=0.5$).  In the near future, the
space density of $z>4$ QSOs predicted from extrapolations of these
models may be usefully compared with the significant numbers of such
QSOs now being identified at these redshifts by the Sloan Digital Sky
Survey (Fan et al.\ 1999).

Inclusion of bright surveys such as the EQS and HEQS do result in
departures from luminosity evolution similar in form to those seen by
other authors (e.g. Hewett et al.\ 1993, La Franca \& Cristaini 1997,
Goldschmidt \& Miller 1998).  The principal region of parameter space
in which such departures are seen lies at $z<0.5$ where those authors
claim a significant flattening of the luminosity function.  The 2QZ
contains relatively few QSOs with $0.35 < z < 0.5$ and does not yet
probe the redshift range $z<0.35$.  Hence there is at present no
inconsistency between the various studies.  Our analysis of the 2dF
survey does however show that, over the redshift range $0.35 < z <
2.3$ and for absolute magnitude $M_B < -23$, pure luminosity evolution
does provide an accurate phenomenological model of QSO evolution.

\section{ACKNOWLEDGMENTS} 

The 2QZ was based on observations made with the Anglo-Australian
Telescope and the UK Schmidt Telescope.  We are indebted to Mike
Hawkins for providing us with colour information on his QSO sample
prior to publication.  NL was supported by a PPARC studentship during 
the course of this work.

\vspace{1.0truecm}
This paper has been produced using the Blackwell Scientific Publications 
\TeX macros.


\begin{thebibliography}{}

\bibitem[Bailey \& Glazebrook 1999]{2dfman}Bailey J., Glazebrook G., 1999, 2dF User Manual, Anglo-Australian Observatory.
\bibitem[Boyle et al. 1988]{bsp}Boyle B.J., Shanks T., Peterson B.A., 1988, MNRAS, 235, 935
\bibitem[Boyle et al. 1990]{bfsp90}Boyle B.J., Fong R., Shanks T., Peterson B.A., 1990, MNRAS, 243, 1
\bibitem[Boyle et al. 1991]{bjsm91}Boyle B.J., Jones L.R., Shanks T., Marano B., Zitelli V., Zamorani G., 1991 in Crampton D., ed., ASP Conf. Ser. 21: The Space Distribution of Quasars, p 191
\bibitem[Boyle \& Terlevich 1998]{bt98}Boyle B.J., Terlevich R.J., 1998, MNRAS, 293, L49
\bibitem[Colless 1998]{colless98}Colless M., 1998, in Mellier Y., Colombi S. eds., Wide Field Surveys in Cosmology, 14th IAP meeting, Editions Frontieres
\bibitem[Cristiani et al. 1995]{hbqs1} Cristiani S. et al.,  1995, AASup, 112, 347
\bibitem[Cristiani \& Vio 1990]{cv90}Cristiani S., Vio R., 1990, AA, 227, 385
\bibitem[Croom 1997]{c97}Croom S.M., 1997, PhD Thesis, University of Durham
\bibitem[Dunlop \& Peacock 1990]{dp90}Dunlop J.S., Peacock J.A., 1990, MNRAS, 247, 19
\bibitem[Efstathiou \& Rees 1988]{er88}Efstathiou G., Rees, M.J., 1988, MNRAS, 230, 5P
\bibitem[Fan et al 1999]{f99}Fan X. et al., 1999, AJ, 118, 1
\bibitem[Green et al. 1986]{gsl86}Green R.F., Schmidt M., Liebert J., 1986, ApJS, 61, 305
\bibitem[Goldschmidt \& Miller 1998]{gm98}Goldschmidt P., Miller L., 1998, MNRAS, 293, 107
\bibitem[Haehnelt \& Rees 1993]{hr93}Haehnelt M.G., Rees M.J., 1993, MNRAS, 263, 168
\bibitem[Hartwick \& Schade 1990]{hs90}Hartwick F.D.A., Schade D., 1990, ARAA, 28, 437
\bibitem[Hawkins \& Veron 1995]{hv95}Hawkins M.R.S., Veron P., 1995, MNRAS, 275, 1102 
\bibitem[Hewett et al. 1993]{hfc93}Hewett P.C., Foltz C.B., Chaffee F.H., 1993, ApJ, 406, L43
\bibitem[Hewett et al. 1995]{lbqs6}Hewett P.C., Foltz C.B., Chaffee F.H., 1995, AJ, 109, 1499
\bibitem[Kochanek 1991]{k91}Kochanek C.S., 1991, ApJ, 379, 51
\bibitem[Koehler et al. 1997]{kgrw97}Koehler T., Groote D., Reimers D., Wisotzki L., 1997, AA, 325, 502
\bibitem[La Franca \& Cristiani 1997]{lc97}La Franca F., Cristiani S., 1997, AJ, 113, 1517
\bibitem[La Franca \& Cristiani 1998]{lc98}La Franca F., Cristiani S., 1998, AJ, 115, 1688, [Erratum to La Franca \& Cristiani 1997]
\bibitem[Lewis et al. 1998]{lgt98}Lewis I.J., Glazebrook K., Taylor K., 1998, in `Fibre optics in Astronomy III', eds S. Arribas, E. Mediavilla, F. Watson, ASP Conf. Ser. 152, p. 71
\bibitem[Marshall et al. 1983]{mtaz83}Marshall H.L., Tananbaum H., Avni Y., Zamorani G., 1983, ApJ, 269, 35
\bibitem[Marshall 1985]{marshall85}Marshall H.L., 1985, ApJ, 299, 109
\bibitem[Meiksin \& Madau 1993]{mm93}Meiksin A., Madau P., 1993, ApJ, 412, 34
\bibitem[Loveday et al. 1998]{sdss98}Loveday J., et al., 1998, in Mellier Y., Colombi S. eds., Wide Field Surveys in Cosmology, 14th IAP meeting, Editions Frontieres 
\bibitem[Page \& Carrera 1999]{pc99}Page J., Carrera F., 1999, MNRAS, in press (astro-ph/9909434)
\bibitem[Peacock 1983]{p83}Peacock J.A., 1983, MNRAS, 202, 615
\bibitem[Peacock et al. 1986]{pml86}Peacock J.A., Miller L., Longair M.S., 1986, MNRAS, 218, 265
\bibitem[Schlegel et al. 1998]{skd98}Schlegel D.J., Finkbeiner D.P., Davis M., 1998, ApJ, 500, 525
\bibitem[Schmidt 1968]{s68}Schmidt M., 1968, ApJ, 151, 393
\bibitem[Smith 1998]{smith} Smith R.J., 1998, Ph.D. Thesis, University of Cambridge.
\bibitem[Smith et al. 1998]{2dfqz98}Smith R.J., Boyle B.J., Shanks T., Croom S.M., Miller L., Read M., 1998, in McLean B.J., Golombek D.A., Hayes J.J.E., Payne H.E., eds., New Horizons from Multi-Wavelength Sky Surveys, Proceedings of the 179th IAU Symposium, Kluwer, p348 
\bibitem[Terlevich \& Boyle 1993]{tb93}Terlevich R.J., Boyle B.J., 1993, MNRAS, 262, 491
\bibitem[Warren et al. 1994]{who94}Warren S.J., Hewett P.C., Osmer P.S., 1994, ApJ, 421, 412

\end{thebibliography}
\end{document}